\documentclass{aa}
\usepackage{graphicx}
\usepackage{txfonts}
\usepackage[figuresright]{rotating}
\begin{document}
\title{Ground-based CCD astrometry with wide field imagers.}
\subtitle{II.   A   star  catalogue  for   M~67:  $WFI@2.2$m~$MPG/ESO$
   astrometry,   $FLAMES@VLT$   radial   velocities.\thanks{Based   on
   observations with the MPG/ESO  2.2m and ESO/VLT telescopes, located
   at  La Silla  and  Paranal Observatory,  Chile,  under DDT  program
   278.D-5027, and the archive material.}$^,$ \thanks{This is paper 33
   of the WIYN Open Cluster Study (WOCS)} }

\author{ R.\ K.\ S.\ Yadav\inst{1,}\thanks{
  present affiliation: 
  Aryabhatta Research Institute of Observational Sciences, 
  Manora Peak, Nainital 263 129, India 
  {\sf email:} rkant@aries.ernet.in}, 
         L.\ R.\ Bedin\inst{2}, 
         G.\ Piotto\inst{1}, 
	 J.\ Anderson\inst{2}, 
	 S.\ Cassisi\inst{3}, 
	 S.\ Villanova\inst{4}, 
   	 I.\ Platais\inst{5}, 
	 L.\ Pasquini\inst{6},  
         Y.\ Momany\inst{7}, and 
	 R.\ Sagar\inst{8}.}
     
  \institute{
  Dipartimento di Astronomia, Universita di Padova, Vicolo dell' Osservatorio 2, I-35122 Padova, EU\\
  \email{[giampaolo.piotto; sandro.villanova]@unipd.it}
  \and
  Space Telescope Science Institute, 3700 San Martin Drive, Baltimore, MD 21218, USA\\ 
  \email{[bedin; jander]@stsci.edu} 
  \and
  INAF-Osservatorio Astronomico di Collurania, via M. Maggini, I-64100 Teramo, EU\\
  \email{cassisi@oa-teramo.inaf.it}
  \and
  Grupo de Astronom\'ia, Departamento de F\'isica, Universidad de Concepci\'on, Casilla 160-C, Concepci\'on, Chile.\\
  \email{svillanova@astro-udec.cl}
  \and
  Dept. of Physics and Astronomy, The Johns Hopkins University, Baltimore, MD 21218, USA\\ 
  \email{imants@pha.jhu.edu}
  \and 
  European Southern Observatory, Garching, Karl-Schwarzschild-Str.\ 2, D-85748, EU\\
  \email{lpasquin@eso.org}
  \and
  INAF-Osservatorio Astronomico di Padova, Vicolo dell'Osservatorio 5, I-35132,  EU\\
  \email{yazan.almomany@oapd.inaf.it}
  \and
  Aryabhatta Research Institute of Observational Sciences, 
  Manora Peak, Nainital 263 129, India \\
  \email{sagar@aries.ernet.in}
 }

\date{Received 13 December 2007 / Accepted 25 February 2008}

\abstract
    {The  solar-age open  cluster  M~67 (C0847$+$120,  NGC~2682) is  a
     touchstone  in studies  of  the old  Galactic  disk. Despite  its
     outstanding role,  the census of  cluster membership for  M~67 at
     fainter    magnitudes    and    their    properties    are    not
     well-established.}
    { Using  the proprietary and  archival ESO data, we  have obtained
      astrometric, photometric,  and radial  velocities of stars  in a
      34$\times$33 arcmin$^2$  field centered on the  old open cluster
      M~67.  }
     {The  two-epoch archival  observations separated  by 4  years and
      acquired  with  the  Wide  Field  Imager  at  the  2.2m  MPG/ESO
      telescope have been reduced with our new astrometric techniques,
      as  described in  the  first  paper of  this  series.  The  same
      observations  served to  derive calibrated  $BVI$  photometry in
      M~67.  Radial  velocities were  measured using the  archival and
      new spectroscopic data obtained at VLT.  }
    {We  have  determined   relative  proper  motions  and  membership
     probabilities  for $\sim$2,400  stars.  The  precision  of proper
     motions  for  optimally   exposed  stars  is  1.9  mas~yr$^{-1}$,
     gradually degrading down to  $\sim$5 mas~yr$^{-1}$ at $V=20$. Our
     relatively precise proper motions at $V>16$ are first obtained in
     this magnitude  range for  M~67.  Radial velocities  are measured
     for  211 stars in  the same  field.  We  also present  a detailed
     comparison  with  recent   theoretical  isochrones  from  several
     independent groups. }
    {For M~67 area we  provide positions, calibrated $BVI$ photometry,
     relative  proper motions,  membership  probabilities, and  radial
     velocities.   We  demonstrate that  the  ground-based CCD  mosaic
     observations just a few years apart are producing proper motions,
     allowing  a  reliable membership  determination.   We produced  a
     catalogue   that  is   made  electronically   available   to  the
     astronomical community.}
    
    \titlerunning{Astrometry of M~67}
    \authorrunning{Yadav et al.} 
    \maketitle

%
%
\section{Introduction}
\label{INTR}
%

Open  clusters  are at  the  basis  of  our understanding  of  stellar
evolution.   Among them,  M~67 (C0847+120)  has a  unique  and special
role.  Due  to their  dynamical evolution and  inevitable dissolution,
old open clusters are rare. Only a handful open clusters are presently
known with ages  comparable to or older than the Sun.   M~67 is one of
them and having a  solar chemical composition, being relatively nearby
($\sim$900~pc) and  with a  low interstellar reddening  is one  of the
most studied open clusters (see Tab.~\ref{fparameters} for fundamental
parameters).   Among  these studies  we  recall  determination of  the
cluster proper-motion  membership (Sanders 1977, Girard  et al.\ 1989,
Zhao  et al.\ 1993),  photometry (Montgomery  et al.\  1993, Sandquist
2004, Momany, Y., et  al.\ 2001, Balaguer-N\'{u}\~{n}ez et al.\ 2007),
high  resolution   spectroscopy  to  determine   chemical  composition
(Tautvai\v{s}iene et al.\ 2000, Randich et al.\ 2006).

In addition,  M~67 harbors an  array of interesting  objects including
blue  stragglers  (Shetrone  \&  Sandquist  2000);  one  AM  Her  star
(Gilliland et al.\ 1991, Pasquini  et al.\ 1994); several X-ray active
stars, detected as coronal sources by ROSAT (Pasquini \& Belloni 1998,
van  den Berg et  al.\ 1999);  a red  straggler and  two sub-subgiants
(Mathieu et al.\ 2003).  Other  studies have been devoted to the white
dwarf  cooling  sequence  (Richer  et  al.  1998),  the  evolution  of
chromospheric  activity along  the RGB  (Dupree et  al.\ 1999)  and in
solar-type stars (Pace \& Pasquini 2004), and the evolution of angular
momentum  of stars  close  to the  turnoff  (Melo et  al.\ 2001).  The
behavior of  light elements  such as lithium  and beryllium  and their
implications for mixing theories have been studied by Pasquini et al.\
(1997), Jones  et al.\ (1999)  and Randich et al.\  (2007).  Extensive
radial velocity surveys have  been carried out, resulting in discovery
of many binaries (e.g., Mathieu et al.\ 1997).  The attempts were made
to  determine  stellar  oscillations  through a  multi  site  accurate
photometry (Gilliland et  al.\ 1991, Stello et al.\  2007).  Given the
uniqueness of this cluster and the similarity of its key properties to
the Sun,  new studies are  under way to explore  spectroscopically the
main sequence stars,  as it has been done in  the other clusters (Pace
et al.\ 2007, Biazzo et al.\ 2007).

Despite  of the  wealth of  accumulated  data on  M~67, a  fundamental
question of  complete cluster membership is still  open.  The existing
high-quality  proper motion  surveys  carried out  by Sanders  (1977),
Girard  et  al.\  (1989),  and  Zhao  et  al.\  (1993)  are  based  on
photographic  plates  limited  to  $V$$\sim$16.  This  is  only  three
magnitudes below the  main-sequence turnoff. However, dedicated and/or
archival  multi-epoch observations with  wide-field CCD  imagers offer
new  opportunities to derive  precise proper  motions within  a short,
few-year time-span,  that are also  deeper by several  magnitudes than
the   photographic  surveys.  The   techniques  of   ground-based  CCD
astrometry (and  photometry) have been developed and  validated on two
nearby globular clusters using  the observations with the MPG/ESO 2.2m
telescope and its wide-field $WFI@2.2$m camera (Anderson et al.\ 2006;
hereafter Paper~I).  With this  setup, just a  2.7 year  time baseline
yields  proper  motions  as  precise as  $\sim$2  mas~yr$^{-1}$.   The
existing  archival  $WFI@2.2$m CCD  frames  for  M~67  made this  open
cluster  a  prime  target  for  deep  proper  motions  based  on  this
techniques. There  is one report on  proper motions in  M~67 for stars
fainter than  $V=16$ by  Lattanzi (1993). No  catalogues of  data ever
followed,  therefore  we  consider  this conference  paper  of  little
contribution  to the  membership studies  in  M~67. In  this study  we
enlarge the number of stars  with known proper motion determination by
a factor of $\sim$2 over a 0.3 deg$^2$ area down to $V\sim21$ which is
5 magnitudes deeper than the existing proper-motion studies of M~67.

In addition,  we analyzed  a large number  of FLAMES$+$\-GIRAFFE\-@VLT
high resolution spectra and determined radial velocities for 211 stars
in the M~67 field good to $\sim$0.5 km s$^{-1}$, the accuracy which is
suitable for reliable RV membership and identification of short-period
binary systems.
All new observational data  are provided to the astronomical community
with  the aim of  convenient access  to these  data for  the follow-up
studies.
 
The main  goal of this study  is to extend the  membership analysis to
much  fainter  magnituds and  present  a  catalogue  of objects  in  a
wide-field  ($\sim$$30'\times30'$) centered  on the  old  open cluster
M~67.   The catalogue provides  new positions,  multi-band photometry,
proper motions, membership probabilities, and radial velocities.

The  structure of  the  paper is  as  follows. In  Sect. \ref{OBS}  we
describe  observations and  data  reduction of  both,  the images  and
spectra.   In Section  \ref{PM} and  \ref{MP} we  present  the cluster
proper motion  membership analysis in two different  ways.  In Section
\ref{CMP} we provide comparison  with other photometric data available
in  the literature.  In  Sect.\ \ref{ISO}  a detailed  comparison with
available models  from different groups is given.   Finally, in Sect.\
\ref{CAT} we present and describe  the electronic catalogue and how to
use it.

\begin{table} 
\caption{Fundamental parameters of M~67. }
\centering
\label{par}
\begin{tabular}{ccc} 
\hline\hline  Parameter & Value & Reference \\ 
\hline
$\alpha$(J2000)  &     08$^h$ 51$^m$ 23.$^s$3 & {\sf this work } \\
$\delta$(J2000)  & +11$^\circ$ 49$'$ 02$''$   & {\sf this work } \\ 
$l$              & 215.$^\circ$688           & {\sf this work } \\
$b$              & 31.$^\circ$923            & {\sf this work } \\ 
$\rm [Fe/H]$     & $+0.03 \pm 0.01$           & {\sf Randich et al.\ (2006)}\\
$E(B-V)  $       & 0.041  & {\sf Taylor (2007) } \\
$(m-M)_{\circ}$  & 9.56-9.72  & {\sf this work } \\
Age[Gyr]         & 3.5-4.8      & {\sf this work } \\
\hline 
\end{tabular} 
\label{fparameters}
\end{table} 

%
%

\begin{table}[h]
\caption[]{Description of the $WFI$ data sets used for this work.}
\label{log}
\begin{tabular}{cccc}
\hline
\hline
Filters      &  Exposure Time & Seeing & Airmass \\
\hline
\multicolumn{4}{c}{ first epoch:\ Feb. 16, 2000}\\
\hline
$B$&9$\times$30s&$\sim$1$''$.0&1.4     \\
$V$&9$\times$20s; 11$\times$30s; 3$\times$120s &$\sim$1$''$.0&1.5     \\
$I$&9$\times$30s&$\sim$1$''$.0&1.4     \\
\hline
\multicolumn{4}{c}{second epoch:\ Feb. 12, 2004} \\
\hline
$V$&1$\times$60s; 1$\times$300s&$\sim$1$''$.2&1.3     \\
\hline
\end{tabular}
\end{table}

\section{Observational Data and Reductions}
\label{OBS}
%

\subsection{Astrometric and Photometric Reductions}

All the  images used in this  study were obtained with  the Wide Field
Imager (hereafter  simply $WFI$) at  the MPG/ESO 2.2m telescope  at La
Silla, Chile.  The $WFI$ consists  of eight 2048$\times$4096  EEV CCDs
with a  pixel scale  of 238 milliarcsec  (mas) which provides  a total
field-of-view  34$\arcmin  \times  33\arcmin$.   More details  on  the
instrumental setup are given in Paper~I.

In order  to calculate  proper motions, we  searched the  MPG/ESO 2.2m
telescope archive for suitable CCD imager frames providing deep images
obtained at separate epochs.  We  found that the open cluster M~67 was
observed in 2000  and 2004 during the technical  nights.  Some details
of the $WFI$ data-set used  in this work are given in Table~\ref{log}.
These  observations  contain short  and  long  exposures, allowing  to
sample  both the  red  giant  branch (RGB)  and  the relatively  faint
main-sequence  (MS) stars.   It should  be noted  that  these archival
datasets have not been taken with the goal of obtaining proper motions
or well-calibrated  photometry.  The second epoch  is represented only
by two images  and individual frames have not  been properly dithered.
As the result, the distribution of registered stars is not contiguous.
Additional CCD  imaging is planned,  however its success  is dependent
upon a proven record of publications.

We  closely followed the  prescription on  reductions of  $WFI$ images
given in  Paper~I. This include standard manipulations  with the pixel
data  such as  de-biasing,  flat-fielding, and  correction for  cosmic
rays. At the core of fitting the star positions and fluxes is the {\em
empirical}  Point  Spread  Function.   In  this concept,  the  PSF  is
represented  by  a  look-up  table   on  a  very  fine  grid.   It  is
well-documented that the shape of PSF changes with position on a chip.
This variability can be captured by  an array of PSFs across the chip.
A fully  automated code is  developed to find appropriate  stars which
can adequately represent the PSF. For practical purposes the number of
PSF stars per chip can vary from 1 to 15, depending on the richness of
a  star-field.  An  iterative process  is  designed to  work from  the
brightest down to  the faintest stars and find  their precise position
and instrumental flux.  A reasonably  bright star can be measured with
a  precision   of  $\sim$0.03   pixel  ($\sim$7~mas)  from   a  single
exposure. In  this step of  empirical PSF fitting we  obtained precise
instrumental magnitudes in $B,V,I$ filters.

Another  issue in  astrometry  with reflectors  is  a large  geometric
distortion in the focal plane that effectively changes the pixel scale
across  the field-of-view.  There  are different  ways to  derive this
geometric distortion. We opted for a $9\times17$ element look-up table
of corrections for each chip, derived from multiple optimally-dithered
observations of the Galactic  Bulge in Baade's Window (Paper~I).  This
look-up table  provides the best  currently available characterization
of geometrical distortions for  the $WFI@2.2$m.  At any given location
on the  detector, a bi-linear  interpolation between the  four closest
grid  points from  a look-up  table provided  the corrections  for the
target point.  The derived look-up  table may have a lower accuracy of
the edges  of a field because  of the way  the self-calibration frames
were dithered  (see Paper~I). An  additional source of  uncertainty is
related to  a possible instability  of distortions for  the $WFI@2.2$m
reported  earlier. This  prompted us  to use  the local-transformation
method to derive proper motions (see Paper~I for details).

In  the local transformation  approach (a  linear adjustment  with six
parameters --  three in each  of the two  coordinates) a small  set of
local reference  stars is  selected around each  target object.  It is
advantageous  to use  pre-selected  cluster members  to  form a  local
reference frame because of  a much lower intrinsic velocity dispersion
among the  cluster members (see  Sect. 2.1.2).  Then,  a least-squares
adjustment is  used to translate  the coordinates from one  frame into
another,   taken  at   different   epochs.  The   residuals  of   this
transformation  are characterizing  relative proper  motions convolved
with  measurement errors.  In  essence, this  is  a classical  ``plate
pair'' method but  extended to all possible combinations  of the first
and second epoch frames. Relative  proper motion of a target object is
an  average of all  displacement measurements  in its  local reference
frame.  The last  step  is  to estimate  the  measurement errors  from
intra-epoch   observations   where   proper   motions  have   a   zero
contribution. A  complete description of  all steps leading  to proper
motions  is  given in  Paper~I.  We did  not  consider  the effect  of
differential  chromatic refraction  because all  CCD frames  have been
obtained within a narrow range of zenith distance (see Sect.\ \ref{ac}
for details).

\begin{figure}
\centering
\includegraphics[width=8.5cm]{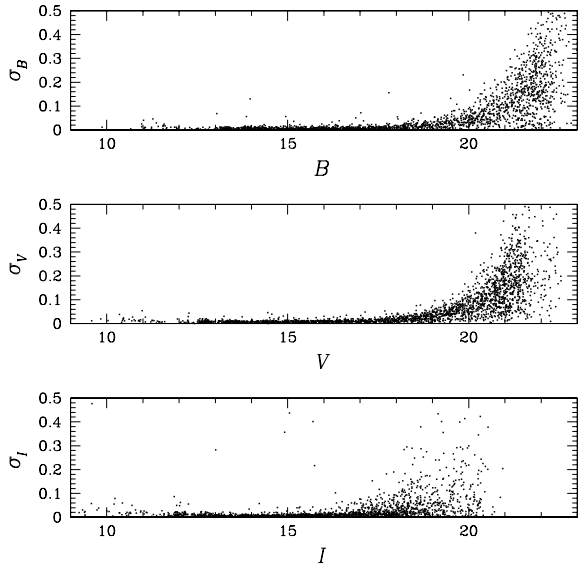}
\caption{Plot of error in magnitudes as a function of magnitude.}
\label{error_mag}
\end{figure}

\subsubsection{Photometric Calibration}

Initially,  we  transformed our  instrumental  magnitudes obtained  in
$B,V,I$  filters  into the  standard  Johnson Kron-Cousins  magnitudes
using                standard                stars                from
Stetson's\footnote{http://cadcwww.hia.nrc.ca/standards/}      secondary
standard star catalogue (Stetson 2000, 2005).
Unfortunately,  the color  and  magnitude intervals  covered by  these
secondary  standards   are  not   ideal  to  properly   calibrate  our
sample. Our  RGB stars are  saturated and the  faint MS stars  are not
present in  the secondary standard  catalogue, thus confining us  to a
narrow  range   of  colors   for  calibrations.   Therefore   for  our
calibrations, we chose the  high-precision $BVI$ photometry of M~67 by
Sandquist (2004).  This photometry  covering the entire color range of
stars  in M~67,  is  calibrated using  the  observations of  Landolt's
standards (Landolt 1992) and, to some extent, also Stetson's secondary
standards.   Sandquist  (2004)  also   made  use  of  photometric  and
spectroscopic  data to  select the  objects most  likely to  be single
stars.  For these  reasons and the ones outlined  in Section \ref{ISO}
we     link     our     photometry     to    that     of     Sandquist
(2004)\footnote{http://\-www.blackwell\-publishing\-.com\-/products\-/journals\-/
suppmat\-/MNR7174\-/MNR7174sm.htm}. This  calibrated photometry covers
only the inner four $WFI@2.2$m CCD chips.  We used the common stars in
dithered  frames between  the pairs  of adjacent  CCD chips  to obtain
individual photometric  zero-points in magnitude. Then,  we solved for
the  global zero-point  and  color  term using  one  of the  following
transformation equations:
$$ B_{\rm std} = B_{\rm ins} + C_b*(B_{\rm ins} - V_{\rm ins}) + Z_b $$
$$ V_{\rm std} = V_{\rm ins} + C_v*(V_{\rm ins} - I_{\rm ins}) + Z_v $$
$$ I_{\rm std} = I_{\rm ins} + C_i*(V_{\rm ins} - I_{\rm ins}) + Z_i $$ 
where $X_{\rm ins}$ is the instrumental magnitude and $X_{\rm std}$ is
the Sandquist's  calibrated magnitude; $C_b,  C_v, C_i$ are  the color
terms  (constants); and $Z_b,  Z_v, Z_i$  are the  global zero-points.
The  quadratic  color  terms  are  negligible.  Our  color  terms  are
consistent     with     those     posted     on     the     $WFI@2.2$m
webpage\footnote{http://www.ls.eso.org/lasilla/sciops/2p2/E2p2M/WFI/zeropoints/}.
A  comparison  of  our  photometry  with  the  Sandquist  and  Stetson
catalogues is presented in Sec.~\ref{CMP}.

In  Figure \ref{error_mag}  we show  our photometric  errors  for each
filter, as function of the corresponding calibrated magnitude.
The photometric  errors (standard  deviation) have been  computed from
multiple observations, all reduced to the common photometric reference
frame in the chosen bandpass.
On  average, the  errors  are  larger than  $\sim$0.03  mag for  stars
fainter than  $V$$\sim$19 mag.  The stars brighter  than $V$$\sim$12.5
mag show higher dispersion because of the image saturation.

\subsubsection{Proper Motions}

We  first  photometrically  selected  likely cluster  members  in  the
color-magnitude diagram and used only these stars as a local reference
frame to transform  the coordinates from one image  into the system of
the other  image at different  epoch and, thus derive  relative proper
motions.   By using predominantly  the cluster  stars, we  ensure that
proper motions will be measured relative to the bulk motion of cluster
stars.   According to Girard  et al.\  (1989), the  intrinsic velocity
dispersion  in M~67  is 0.2  mas~yr$^{-1}$. Over  the  four-year epoch
difference that would  result in a displacement of  only 0.8 mas which
is  by   a  factor   of  10  smaller   than  the   random  measurement
errors. Conversely, the tangential  velocity dispersion of field stars
is  by  a factor  of  $\sim40$  larger  than the  cluster's  intrinsic
velocity  dispersion.  For  field  stars, proper  motions are  clearly
dominating over the measurement errors  and this has an adverse effect
on the coordinate transformations.   We iteratively removed some stars
from the  preliminary photometric member list that  had proper motions
clearly inconsistent with cluster membership, even though their colors
placed them near the fiducial cluster sequence.

In order  to minimize  the influence of  any unaccounted  for residual
distortion  and refraction effects  on proper  motions, we  used local
transformations  based  on   the  nearest  $\sim$20  reference  stars,
typically  extending over $\sim$300  arcsec.  These  are well-measured
cluster stars of any magnitude selected  on the same CCD chip, as long
as  their  preliminary  proper   motion  is  consistent  with  cluster
membership.  No systematics larger  than our random errors are visible
close to the corners or edges of chips.
In  order to  avoid  possible filter-dependent  systematic errors,  we
measured  proper motions  in  the  $V$ bandpass  only,  for which  the
geometrical distortion corrections  were originally derived (Paper~I).
Individual  errors  of  proper  motions  for single  stars  have  been
estimated as described in Section 7.3 of Paper~I.

In  a  nutshell,  we  estimate  the intra-epoch  rms  error  from  all
first-epoch plates locally transformed  into the same reference frame.
Due to  the lack  of adequate statistics  (too few images)  we assumed
that  the  second  epoch frames  have  the  same  errors for  a  given
instrumental magnitude.  The proper  motions errors have been computed
as the  rms of the proper  motion, obtained from  solving locally each
first-epoch frame into second-epoch  frame. These errors, however, are
not totally  independent because the same  frame is used  for a second
epoch.  Therefore,  to obtain our  best estimate of the  proper motion
standard error we added in  quadrature initial proper motion error and
the adopted error of the second epoch image.

\begin{figure}[h]
\centering
\includegraphics[width=8.5cm]{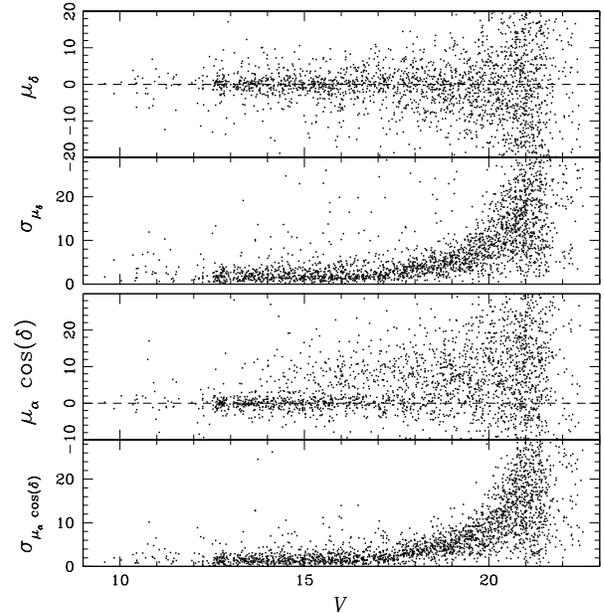}
\caption{Plots of proper motions  and their standard deviations versus
         visual magnitude in mas~yr$^{-1}$.}
\label{error_pm}
\end{figure}

Figure  \ref{error_pm}  shows  the  distribution  of  standard  errors
($\sigma$) in proper motions for both coordinates as a function of $V$
magnitude.   The  precision  of   our  proper  motion  measurement  is
$\lesssim$ 0.05  $WFI$ pixels in 4  yrs down to  $V\sim$18 mag (i.e.,\
$\sigma$$\lesssim$ 3 mas~yr$^{-1}$).  At fainter magnitudes the errors
gradually  increase,  reaching $\sim$6  mas~yr$^{-1}$  at $V$=20.  The
stars brighter than $V\sim$ 12.5  mag show a higher dispersion because
of the image saturation.

\subsubsection{Astrometric Calibration}
\label{ac}

One of  our goals  is to provide  in the  area of M~67  an astrometric
frame  on the  International Celestial  Reference System  (ICRS).  Our
astrometric  reductions  leads  us  to the  distortion-free  cartesian
coordinates of all objects.  The positions of stars in each frame have
been  corrected for geometric  distortion using  a look-up  table from
Paper~I,  and then  averaged into  a  common reference  frame using  a
six-parameter linear transformation.   The astrometric {\it positions}
of  this  catalogue  are  still  affected  by  differential  chromatic
refraction  (DCR) effects  and  the limitations  on  the stability  of
derived  geometrical  distortions,  as  described  in  Sect.\  5.2  of
Paper~I.

To  link  the  present M~67  catalogue  to  ICRS,  we used  the  UCAC2
catalogue  (Zacharias et  al.\ 2004)  as  a reference  frame.  It  was
trimmed down  to the stars  with positional accuracies better  than 75
mas.   Owing to  the  excellent geometric  distortion  solution and  a
reasonable stability  of the intra-chip  positions it was  possible to
apply a single plate model  involving linear and quadratic terms and a
small but significant cubic term in each coordinate.
This solution also absorbes  effects caused by differential refraction
but not its chromatic component.
The  standard error  of equatorial  solution is  only 30  mas  in each
coordinate,  thus  indicating  a   great  potential  of  the  WFI  for
astrometry.

A  total of  489  reference stars  were  used in  this solution.   The
provided J2000  coordinates in Table ~\ref{catalogT} are  on the epoch
2000.13.

A  compilation of  kinematic and  photometric data  from a  variety of
sources (Platais, unpublished) allowed us to pre-select a fairly clean
list  of cluster members  down to  $V\sim16$, which  then was  used to
derive a better geometrical center of M~67. The marginal distributions
of cluster members along right ascension and declination were fit with
a  Gaussian. This  fit yielded  a  cluster center  listed in  Table~1,
accurate to about $3\arcsec$.

The  proper  motions  in  this   work  have  not  been  corrected  for
differential chromatic  refraction effects (DCRE) as  done in Paper~I.
Nonetheless, we verified  the differential chromatic refraction effect
and  found it to  be negligible  in this  data set  owing to  the very
similar observing conditions at both epochs.

The only determination of absolute proper motion for M~67 is that from
using two  Hipparcos stars (HIP  43465, 43491) -- apparent  members of
the  cluster (Baumgardt  et al.\  2000). According  to this  study the
absolute proper motion  for M~67 is $\mu_{\alpha} \cos  \delta = -6.5$
and $\mu_{\delta}  = -6.3$  mas~yr$^{-1}$.  Although a  few background
galaxies are visible in our images, they are not suitable for deriving
the correction needed to convert  our relative proper motions into the
absolute proper motions.

%
%
\subsection{Radial Velocities: Reduction and Calibration}
\label{rv}
%

With the aim of  obtaining additional membership information for stars
in   the    region   of   M~67,   we   searched    the   ESO   archive
(http://archive.eso.org/) for sets of FLAMES\-+\-GIRAFFE\-@VLT (MEDUSA
mode)  observations taken simultaneously  with calibration  lamp.  The
MEDUSA  mode  allows  obtaining  about  130 spectra  at  once  over  a
field-of-view  of  $25$ arcmin  in  diameter.   We found  observations
performed with three different set-ups (HR02, HR14, HR15).

In addition to these data sets,  we used the data (HR15N) obtained for
a different project\ in DDT time (278.D-5027).  They were collected to
determine  bona  fide  possible  single solar-type  members  in  M~67,
closely resembling the  Sun. The aim of this program  is to find solar
analogues among the M~67 solar  type stars.  Full analysis of the data
is  in progress  and will  be reported  elsewhere (Biazzo  et  al., in
preparation).

The instrumental set-up for each  run is given in Table\ \ref{HRconf},
indicating the wavelength range, resolution, the number of plates, and
the date of observations.

\begin{table}[t]
\caption{Radial velocity observations.}
\centering
\begin{tabular}{l c c c c}
\hline\hline
Set-up & $\Delta \lambda$ (\AA)& R & plates & OBS-DATE \\
\hline
HR02  & 3854-4049 & 19600 & 2 & 19/12/2002 \\
HR14  & 6383-6626 & 28800 & 5 & 26-28/02/03 \\
HR15  & 6607-6965 & 19300 & 6 & 26-28/02/03 \\
HRN15 & 6470-6790 & 17000 & 3 & 06-11-23/02/07 \\ 
\hline
\end{tabular}
\label{HRconf}
\end{table}

%
    \begin{figure}[ht!]
    \centering
    \includegraphics[width=9.0cm]{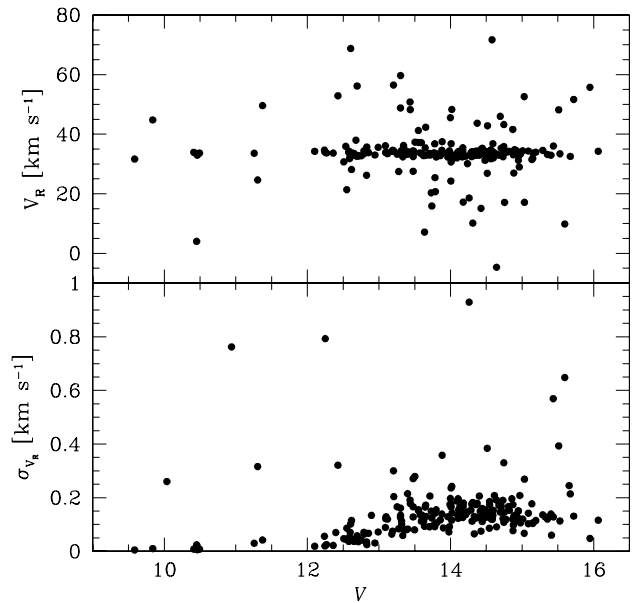}
       \caption{Heliocentric radial  velocities (top panel)  and their
pipeline-errors (bottom panel) of stars from the spectroscopic sample,
plotted as a function of $V$ magnitude.}
          \label{vr1}
    \end{figure}

%
%

The  spectroscopic  data  were  reduced  using  the  GIRAFFE  pipeline
GIRBLDRS\footnote{Blecha     et     al.\     (2000).      See     {\sf
http://\-girbldrs.sourceforge.net/},  for  GIRAFFE pipeline,  software
and  documentation.},  in  which  the  spectra  have  been  de-biased,
flat-field corrected, extracted, and wavelength-calibrated, using both
prior and simultaneous calibration-lamp spectra.

We also used the {\sf  gyCrossC.py} utility of the GIRAFFE pipeline to
measure heliocentric radial velocity  ($\rm V_R$). For each set-up, we
used an appropriate K0 type  template spectra developed by the GIRAFFE
pipeline team.

The formal errors  reported in the catalogue are  merely the output of
pipeline   which   is  clearly   underestimating   the  true   errors.
Nevertheless these estimates retain the information on the goodness of
the fit of cross-correlation function.

There  are  signs of  systematic  errors  in  $\rm V_R$,  if  repeated
measurements of radial velocity from different set-ups are considered.
To estimate the amount of such  errors, we computed the r.m.s. for all
determinations  of  $\rm V_R$  of  the  same  star and  averaged  that
estimate  for all  multiple measurements  of radial  velocity.  As the
result, a single  measurement of radial velocity could  be affected by
systematics reaching up to $\sim$0.6  km s$^{-1}$. Perhaps, this is an
upper  limit of  systematic  offsets  because some  stars  may have  a
physical  explanation of $\rm  V_R$ variations.  To estimate  the real
uncertainty in radial  velocities, we suggest to sum  up in quadrature
this estimate and the formal fitting error.

The heliocentric  radial velocities and relative errors,  are shown in
Fig.~\ref{vr1} as a  function of $V$ magnitude.  Several  stars have a
radial velocity very different  from the cluster's mean velocity. This
can be due to the  field stars contaminating the spectroscopic sample,
or  short-period spectroscopic  binaries which  are bona  fide cluster
members.  We determined  the median $\rm V_R$ of  our sample of radial
velocities and  then estimated the dispersion around  this median $\rm
V_R$, assuming  a Gaussian  distribution and considering  the 68$^{\rm
th}$  percentile  from  the   median.   Then  we  iteratively  used  a
3.5-$\sigma$  clipping  to  achieve  a  convergence.  The  final  mean
heliocentric  radial  velocity for  M~67  estimated  from 141  cluster
members is $\overline{V_{R}}=33.67\pm0.09\ $ km s$^{-1}$.

As indicated by Tab.\ \ref{VRtab}, our value of $\overline{\rm V_{R}}$
is consistent with those derived by the other authors.

\begin{table}[h]
\caption[]{Mean weighted  radial velocities for M~67  derived from the
data in {\sf WEBDA}.}
\label{r_velocity}
\begin{tabular}{ccl}
\hline
\hline
$\overline{\rm V_{R}}$ [km $s^{-1}$] & $N_s$    & Reference\\
\hline
34.07 $\pm$ 0.10 & 144              & Mathieu et al. (1986)   \\
34.77 $\pm$ 0.21 &  50              & Glushkova et al. (1990) \\
33.54 $\pm$ 0.13 &  27              & Glushkova, unpublished  \\
32.89 $\pm$ 0.18 &  10              & Shetrone et al. (2000)  \\
33.85 $\pm$ 0.16 &  26              & Melo et al. (2001)      \\
33.67 $\pm$ 0.09  & 141              & this work               \\ 
\hline
\hline
\label{VRtab}
\end{tabular}
\end{table}

\section{Cluster CMD decontamination}
\label{PM}

In this  section we probe the  effectiveness of our  proper motions in
separating clusters  stars from the  field stars using  a vector-point
diagram (VPD) in combination with a color-magnitude diagram (CMD).

\begin{figure*}[!ht]
\centering
\includegraphics[width=\textwidth]{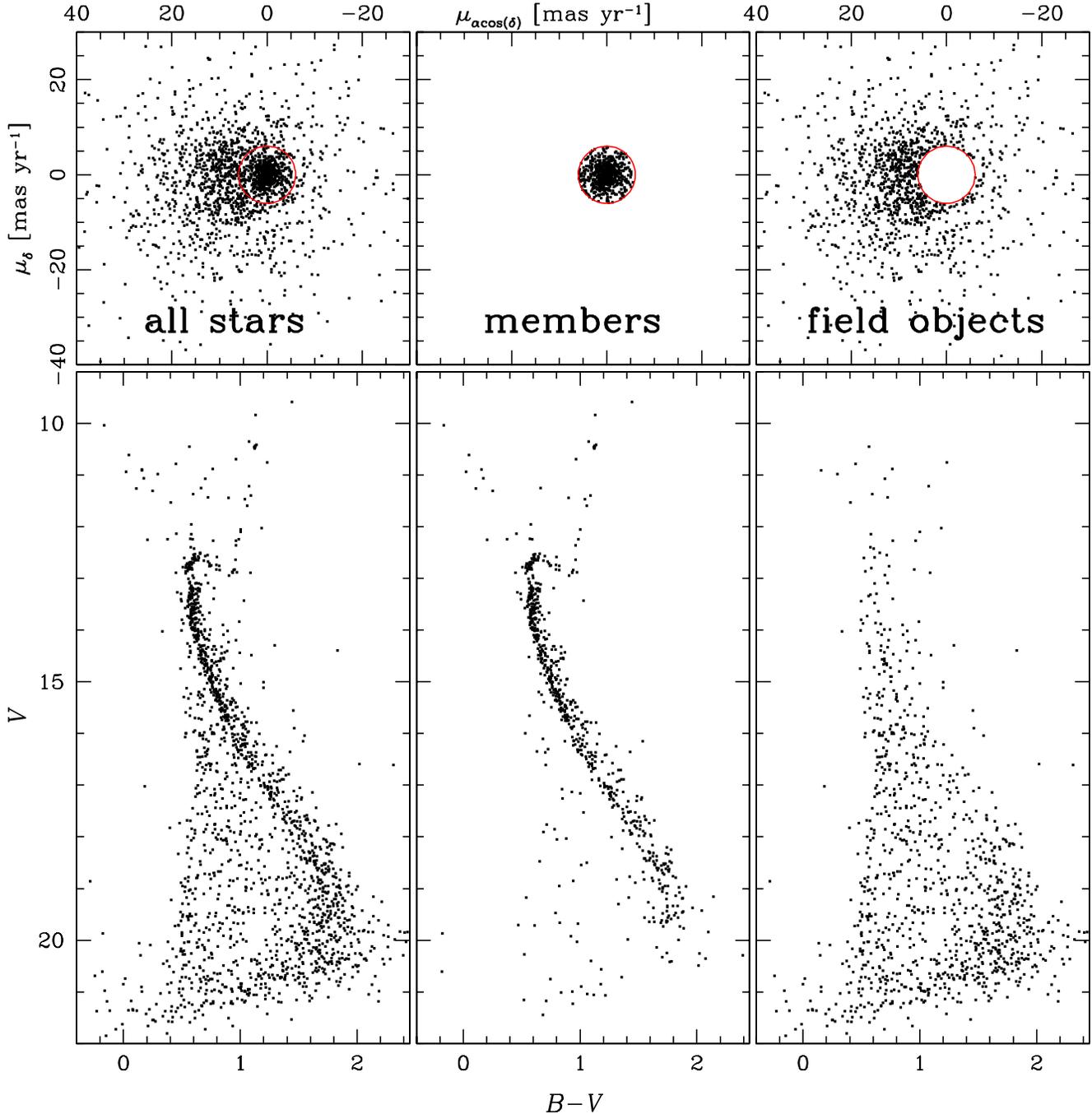}
\caption{
	{\em (Top panels)} Proper motion vector-point diagram.  
	Zero point in VPD is the mean motion of cluster stars.
	{\em (Bottom panels)} Calibrated $(B-V)$, $V$ color-magnitude
	diagram.
	{\em (Left)} The entire sample; 
	{\em (Center)} stars in VPD with proper motions within 6 mas~yr$^{-1}$
         around the cluster mean.
	{\em (right)} Probable background/foreground field stars in the area
	of M~67 studied in this paper.
	All plots show only stars with proper motion
	$\sigma$ smaller than $\sim$20 mas~yr$^{-1}$ in each
	coordinate.  }
\label{cmd_inst}
\end{figure*}

For stars  with $BV$ photometry,  in Fig.\ \ref{cmd_inst} we  show the
vector-point  diagram (top  panels), and  the CMDs  in  the $(B-V),~V$
plane (bottom panels).
In the  left panels  there is  the whole sample  of stars;  the middle
panels display  what we consider  to be probable cluster  members; the
right panels show predominantly the field stars.

In vector-point diagrams around the  cluster centroid we draw a circle
with  the  radius of  6  mas~yr$^{-1}$.  Provisionally,  we define  as
cluster  members all  points in  VPD within  this circle.   The chosen
radius  is a  compromise  between loosing  cluster  members with  poor
proper motions  and including field  stars sharing the  cluster's mean
proper motion. Even  this crude division into cluster  and field stars
is demonstrating the power of proper motions derived in this study.  A
formal  determination of  membership  probability is  given in  Sect.\
\ref{MP}.

\begin{figure*}[!ht]
\centering
\includegraphics[width=\textwidth]{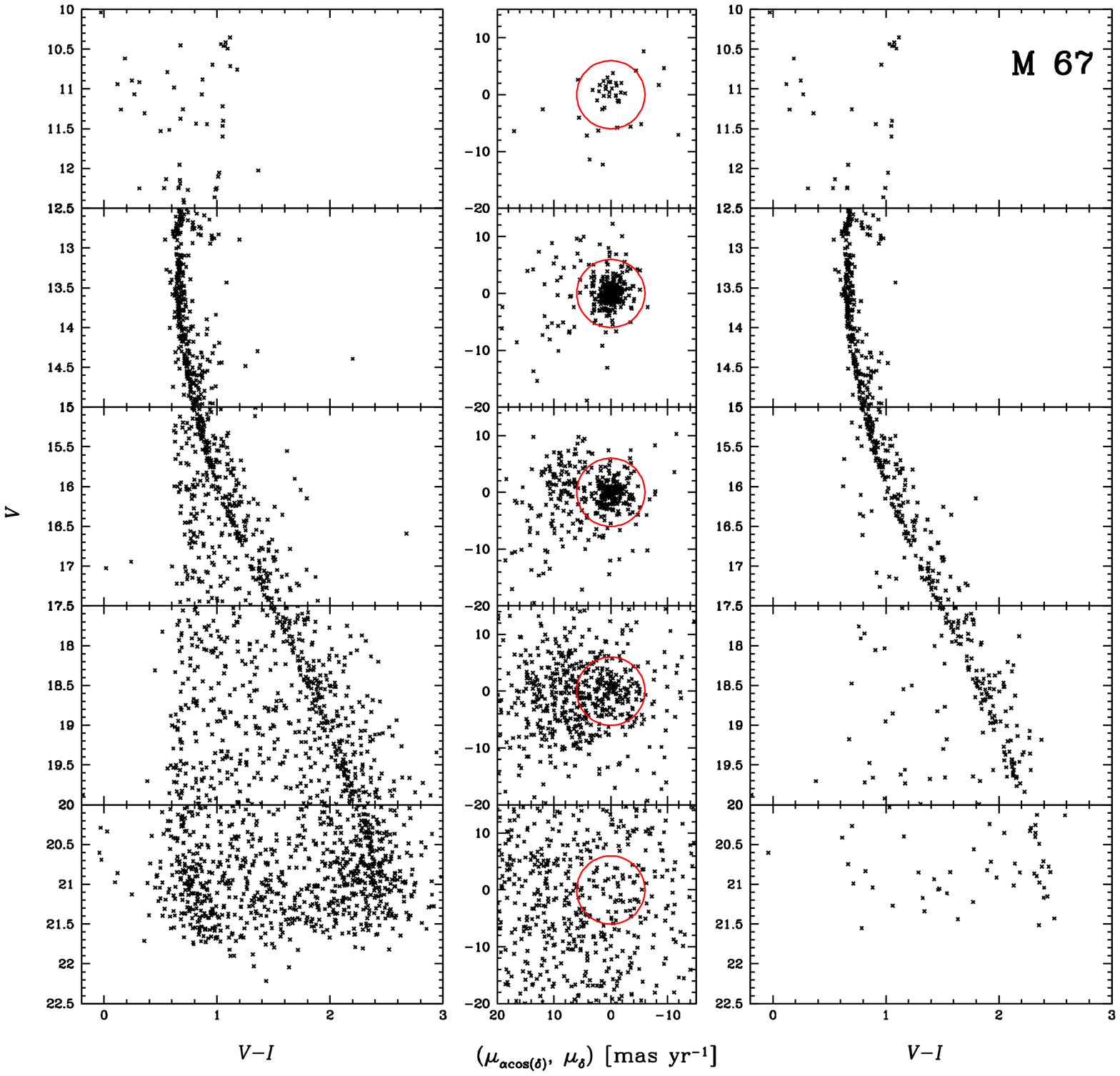}
\caption{
	{\em (Left:)} Color-magnitude diagram for stars with proper
        motions.
 	{\em (Middle:)} Vector-point diagram for the same stars in
	corresponding magnitude intervals. A circle in each plot
	shows the adopted membership criterion at 6 mas~yr$^{-1}$.
 	{\em (Right:)}  Color-magnitude diagram for stars assumed
	to be cluster members.  
}
\label{cmd_II}
\end{figure*}

As expected, the fainter stars  have less well measured proper motions
because     of    rapidly     decreasing     signal-to-noise    ratio.
Figure~\ref{cmd_II}  illustrates  this fact  in  the $(V-I),~I$  plane
binned  along  the  magnitude   axis.   We  note  that  an  acceptable
membership    is   established   even    for   saturated    stars   at
$V\lesssim$12.5. Note that photometry for bright stars is derived only
from short  exposures and then adjusted via  the zero-point correction
to photometry from long exposures.

%
%

%
%

The separation between  the cluster members and field  stars is not as
obvious as it  is for the globular clusters  presented in Paper~I.  In
the case of M~67 the distributions of cluster and field proper motions
have    a   substantial   overlap,    separated   only    by   $\sim$8
mas~y$^{-1}$.  Therefore  it  is  not  surprising that  at  the  faint
magnitudes  where  proper  motion  errors are  large,  the  separation
between the cluster and field is poor.

Finally, we note  that our field star CMD (right  lower panel in Fig.\
\ref{cmd_inst})  shows no  trace of  the anomalous  clump  detected by
Balaguer-N\'{u}\~{n}ez et  al.\ (2007) in  in the $V,b-y$  and $V,v-y$
color-magnitudes diagrams  .  Perhaps, this is  not surprising because
these  authors cover a  larger area  around M~67  than our  study and,
according to  Balaguer-N\'{u}\~{n}ez et al.,  the stars in  this clump
mysteriously avoid the central parts of M~67.
 
%
\section{Determination of Membership Probability}
\label{MP}
%

The vector-point diagram (Fig.\  \ref{cmd_inst}) of our proper motions
clearly shows  two populations:\ a  tight clump at  $\mu_{\alpha} \cos
\delta$ $=$  $\mu_{\delta}$ $=$ 0.0  mas~yr$^{-1}$ mainly representing
the  cluster stars  and  an overlapping  broad  distribution of  field
stars.  Vasilevskis et al. (1958)  were the first one to formulate the
proper-motion membership probability as

\begin{equation}
P_\mu=\frac{\Phi_c}{\Phi_c+\Phi_f},
\end{equation}

\noindent  where $\Phi_c$  is the  distribution of  cluster  stars and
$\Phi_f$  --  the distribution  of  field  stars  in the  vector-point
diagram.  Usually,  these distributions are  represented by elliptical
normal  frequency  distributions, i.e.,  Gaussians,  although in  some
special cases  the field star  distribution can locally be  adopted as
non-Gaussian (Jones  1997, Platais et  al.\ 2001). There  are numerous
papers  discussing various  aspects of  calculating the  parameters of
these  distributions and  membership probabilities  in  general, e.g.,
Jones  (1997), Girard  et  al.\ 1989,  Balaguer-N\'{u}\~{n}ez et  al.\
(2007) and    references   therein.    For    membership   probability
calculations, we have chosen the local sample method, which is equally
reliable in rich open clusters  such as NGC~188 (Platais et al.\ 2003)
as well as in sparse clusters  like IC~2391 (Platais et al.\ 2007). In
this  method, a local  sample of  stars is  selected to  represent the
properties of a target star as close as possible.  It assures a smooth
transition  of calculated  membership probabilities  as a  function of
magnitude and also accounts for the changes in the magnitude dependent
cluster-to-field  star ratio.  It  has been  recognized that  a star's
brightness and spatial location are  the two critical parameters to be
considered in the membership  probability calculations (Girard et al.\
1989,   Jones  1997).    To   counter  the   effect   of  a   changing
cluster-to-field star ratio, we chose a sliding 2.5-mag-wide magnitude
bin centered onto  the target.  The exception are  the extremes of the
magnitude range where a magnitude bin is fixed at 2.5 mag. The spatial
distribution was ignored due to the relatively small size of our field
($\sim$$30\arcmin\times30\arcmin$) with  respect to the  radial extent
of M~67 out to at  least one degree (Sanders 1977).  Another component
of our approach is to  parameterize the magnitude dependence of proper
motion  dispersions for  cluster and  field stars.   That can  be done
using  photometrically selected  samples  of cluster  and field  stars
(Fig.\   \ref{cmd_II}).   Finally,   the   center   of   proper-motion
distribution  for  cluster stars  was  adopted  at $\mu_{\alpha}  \cos
\delta$ $=$  $\mu_{\delta}$ $=$ 0.0 mas~yr$^{-1}$ but  for field stars
it  was estimated from  the photometrically  selected field  stars and
also parameterized as a function of magnitude. It should be noted that
in  $\mu_{\alpha} \cos  \delta$-coordinate  the center  of field  star
distribution  in  the VPD  is  at  $\sim$$+6$  mas~yr$^{-1}$ which  is
significantly larger than that reported  by Girard et al.\ (1989).  We
made   no  effort   to  account   for  the   individual  proper-motion
uncertainties  other  than   the  parameterizations  mentioned  above.
Poorly  measured  proper motions,  i.e,  with uncertainties  exceeding
$\sim$10  mas~yr$^{-1}$  are   insufficient  for  reliable  membership
determination in M~67. Nevertheless, we  kept these stars for the sake
of  demonstration the  achieved accuracies  over the  entire magnitude
range.

\begin{figure}[!ht]
\centering
\includegraphics[width=8.5cm]{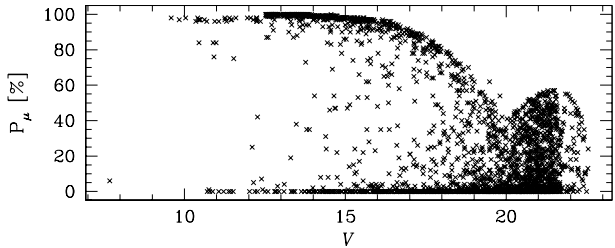}
\caption{Membership  Probabilities as function  of the  $V$ magnitude,
for all the stars in our catalogue.}
\label{VvsMP}
\end{figure}

Calculation  of  the two  remaining  parameters  (heights of  Gaussian
peaks)   needed  to   fully   characterize  the   cluster  and   field
distributions in the VPD is straightforward (Platais et al.\ 2003). We
note  that  the distribution  of  field  stars  appears to  be  nearly
circular that further simplifies the calculations. The distribution of
membership  probabilities (Fig.\ \ref{VvsMP})  in effect  reflects the
degree  of reliability  of our  proper motions.  Thus,  the separation
between the cluster and field  is convincing for $V<18$.  In the range
$18<V<20$ the decreasing maximum  $P_\mu$ indicates a steadily growing
contamination by  field stars. The unexpected rise  of maximum $P_\mu$
at $V>20$  is an artifact  caused by a  nearly total blending  of both
cluster and field proper motion distributions. In other words, in this
magnitude  range the  accuracy  of proper  motions  is inadequate  for
reliable calculations of membership probability.

%
\section {Comparison with existing photometric data}
\label{CMP}
%

\begin{figure*}[ht!]
\centering
\includegraphics[width=\textwidth]{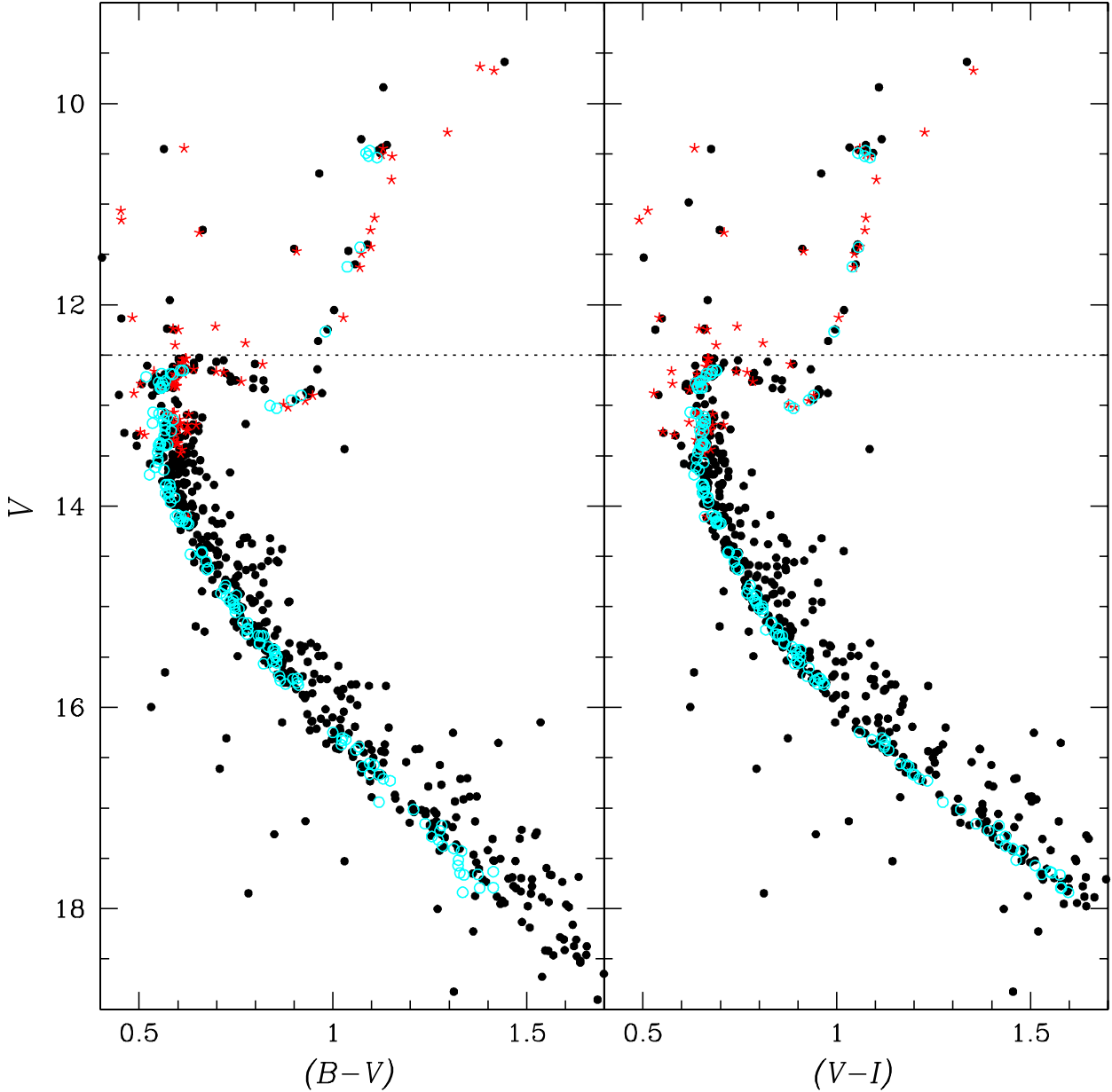}
\caption{Comparison of our photometry (filled circles) with that
from Stetson (2000, 2005; marked by asterisks) and from Sandquist (2004;
marked by the open circles). }
\label{comp}
\end{figure*}

In  Figure   \ref{comp},  we  show  color-magnitude   diagram  in  the
[$(B-V),V$], and [$(V-I),I$] planes for  the stars of our sample which
have a membership probability $P_{\mu} > 60\%$ (see Sect.~\ref{MP} for
details).
The saturation level of deep  $V$ exposures is indicated by the dotted
line.
The  stars from  the secondary  standard catalogue  by  Stetson (2000,
2005) are over-plotted as red asterisks.  The cyan color-coded circles
show the location  of the Sandquist (2004) single  stars. As expected,
the  location  of  main  sequence  from our  photometry  and  that  by
Sandquist is  nearly identical  because our photometry  was calibrated
using Sandquist's  data. The upper  main sequence and giant  branch in
[$(B-V),V$]  have  some  small   offsets,  if  different  sources  are
considered. In part,  that can be explained by  retaining single stars
only in the Sandquist photometry.

\begin{figure}[ht!]
\centering
\includegraphics[width=8.5cm]{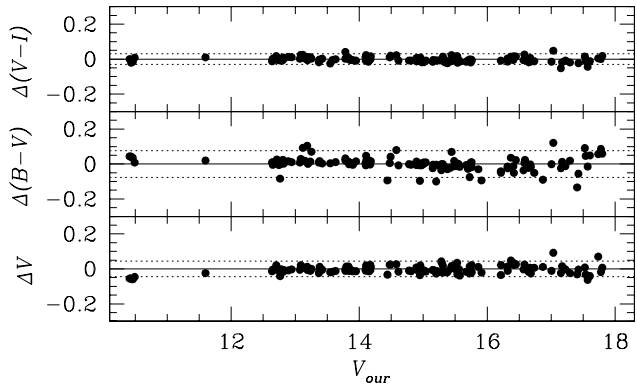}
\caption{Star-by-star comparison of our photometry with that
from Sandquist (2004).}
\label{comp_san}
\end{figure}

\begin{figure}[ht!]
\centering
\includegraphics[width=8.5cm]{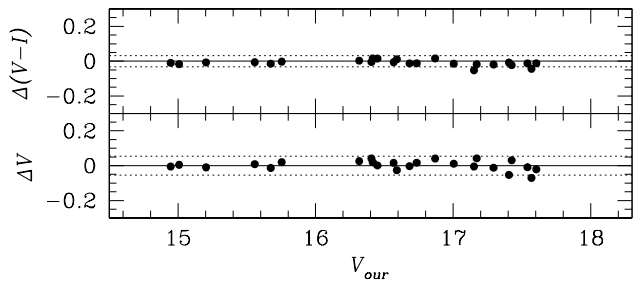}
\caption{Comparison of our photometry with the one by Taylor (2008,
in press). }
\label{comp_tay}
\end{figure}

In Fig.~\ref{comp_san} we show the photometric residuals as a function
of magnitude and color for  the common stars between Sandquist and our
data.
Sandquist   (2004)   presented   extensive  comparisons   with   other
photometric data  from the literature, in  particular, with Montgomery
et  al.\ (1993).   We  refer the  reader  to this  study for  detailed
comparisons.
The  referee suggested  to make  a request  for the  Ben  Taylor's new
improved $VRI$  photometry and perform  a direct comparison  with this
new catalogue.   Ben Taylor  kindly released his  photmetry to  us and
this  comparison  is  shown   in  Fig.~\ref{comp_tay},  in  the  sense
`ours-Taylor'  photometry. If  there are  any offsets,  then certainly
less than 0.01 mag in both, $V$ magnitude and $V-I$ color.
%

%
\section {Comparison with theoretical isochrones}
\label{ISO}
%

\begin{figure*}
\centering
\includegraphics[width=\textwidth]{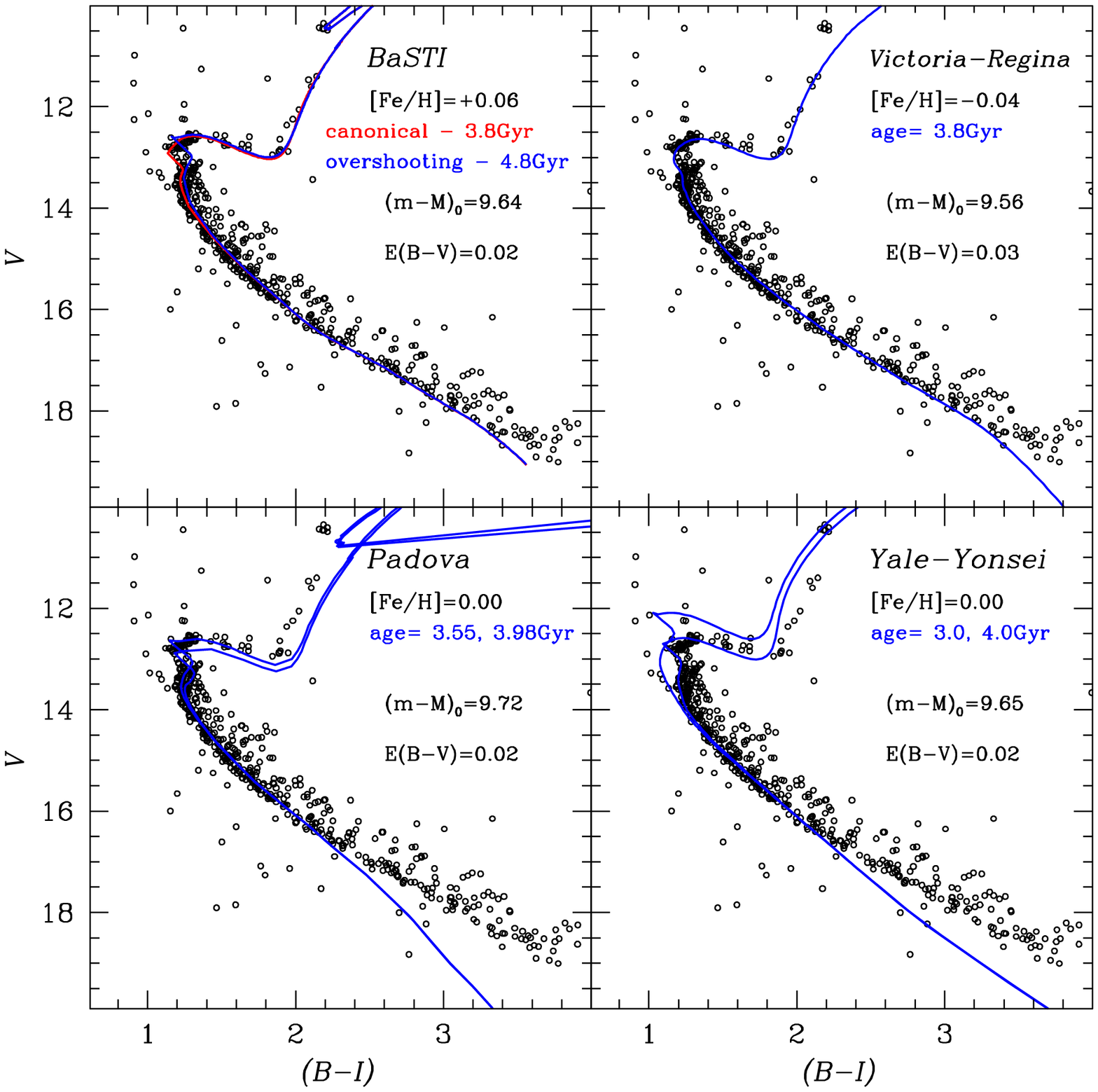}
\caption{
Comparison  of our  empirical  data with  theoretical isochrones  from
independent  libraries  of  stellar  models  (see the  text  for  more
details). The  fit between theory and observations  has been performed
by  adopting the  labeled values  for  the true  distance modulus  and
reddening.}
\label{cmd_teo}
\end{figure*}

CMDs  of Galactic  open clusters  have long  been used  as a  tests of
stellar models  for intermediate-mass stars. The  most important issue
is related to  the long-standing problem of the  real extension of the
convective core during the central H-burning stage.
How much larger  the convective core extension is  with respect to the
canonical   prediction  provided   by   the  classical   Schwarzschild
criterion?    [i.e.\  which   is   the  amount   of  convective   core
overshooting,  see   e.g.\  Sandquist  (2004),   Cassisi  (2002),  and
references therein].

Open cluster studies have shown that stellar models must allow for the
occurrence  of convective  core  overshooting in  order  to provide  a
satisfactory match to the observed  CMDs (see Kalirai et al.\ 2001 and
references therein).
On  the  other  hand,  a  still  unsettled issue  is  related  to  how
convective core  overshooting has to  decrease in real stars  when the
size of  the canonical convective  core decreases with  the decreasing
total stellar mass (e.g., Woo \& Demarque 2001).
In the  majority of  Galactic open clusters  this is not  an important
problem. However, an age  difference exceeding $\pm 1$Gyr with respect
to the age  of M~67 is sufficient for stars  to enter the evolutionary
phase with a very extended  convective core, even encroaching the mass
range normally having a radiative core.
In the  case of M~67, the mass  of stars at the  main sequence turnoff
(TO)  is $\sim$1.2M$_\odot$, which  is in  the critical  mass interval
where the amount of  overshooting is decreasing with mass.  Therefore,
M~67  is  crucial for  exploring  the  efficiency  of convective  core
overshoot in  stars whose  mass is large  enough to have  a convective
core  but not  so large  to  allow the  presence of  a quite  extended
convective core.

For this  reason, M~67 has been  the subject of  many detailed studies
such as  those recently performed  by Sandquist (2004),  Vandenberg \&
Stetson (2004), and Vandenberg et al. (2007).
The work of  VandenBerg \& Stetson (2004) was  specifically devoted to
constraint  the efficiency  of convective  core overshooting  in stars
whose mass  is similar  to that of  the stars  currently at the  TO of
M~67.  Their main conclusion --also  supported by the results shown by
Pietrinferni et  al.\ (2004)-- is  that the extension  of overshooting
region  has  to  be very  small  but  not  zero  for masses  of  about
$1.2M_\odot$, at least for solar metallicity.

A detailed  discussion of this  issue is clearly  out of the  scope of
this paper.  Nonetheless, a comparison  of our observational  data for
M~67 and  the theoretical isochrones available  in various independent
libraries of stellar models  should provide further insights about the
capability  of commonly-used  sets  of stellar  models in  reproducing
empirical constraints by  CMD of the old Galactic  open clusters.  For
this purpose,  we adopted the  following sets of  theoretical tracks:\
the  BaSTI models  provided by  Pietrinferni et  al.\ (2004,  see also
Cordier et al.\ 2007); the  Padova models published by Girardi et al.\
(2000); the Yonsei-Yale  isochrones provided by  Yi et al.\  (2001) in
their  second release;  and  the Victoria-Regina  models published  by
VandenBerg,  Bergbusch   \&  Dowler  (2006).   In  the   case  of  the
Yonsei-Yale isochrones,  we apply their set  of isochrones transformed
into  the   observational  plane  using   the  $T_{\rm  eff}$-to-color
transformation tables provided by Green, Demarque \& King (1987).

The  comparison  among  various  fits  of  theoretical  isochrones  to
observational CMDs is shown in Fig.\ \ref{cmd_teo}.  Before discussing
this figure, it  is necessary to briefly review  the best estimates of
metallicity, reddening and distance  of M~67.  The metallicity of M~67
is essentially the  solar or maybe slightly lower  (Taylor 2007). This
result   is   well-established   on    the   basis   of   both   high-
(Tautvai\v{s}iene et al.\ 2000) and low-resolution spectroscopy (Friel
\& Janes 1993).  The  existing estimates of interstellar extinction in
the direction of M~67 have been recently re-analyzed by Taylor (2007).
The most likely mean value of $E(B-V)=0.041\pm0.004$ derived by Taylor
is slightly larger than the value suggested by Schlegel et al.\ (1998)
and  those provided  by the  fits between  theoretical  isochrones and
empirical  data (e.g, VandenBerg  \& Clem  2003). Given  the intrinsic
uncertainties in  calibration of $E(B-V)$, the true  reddening of M~67
could be anywhere between 0.02 and 0.04 mag.

As for the  distance to M~67, Percival \&  Salaris (2003) measured the
distance modulus at  $(m - M)_\circ = 9.60 \pm  0.09$, adopting $E(B -
V) = 0.04 \pm 0.02$ and [Fe/H] $=$ $0.02 \pm 0.06$.
Their  measurement  has  been  obtained  by  applying  the  MS-fitting
technique to  the CMD from Montgomery  et al.\ (1993) data  by using a
sample of local G and  K dwarfs with accurate Hipparcos parallaxes and
a near-solar  metallicity.  The Percival \&  Salaris's (2003) estimate
was subsequently confirmed by  Sandquist (2004), who obtained the same
distance modulus using new photometric data.
 
The  comparison shown  in Fig.\  \ref{cmd_teo} is  quite informative:\
within  the errors the  BaSTI (Pietrinferni  et al.\  2004) isochrones
with [Fe/H]  $=$0.06 reproduce nicely the  CMD with $E(B -  V) = 0.02$
and $(m - M)_\circ = 9.64$, in agreement with previous determinations.
Both isochrones with and  without convective core overshoot are shown.
The adopted metallicity appears to  be the upper limit of any existing
estimate of [Fe/H] in M~67.
The MS and the red giant  branch appear to match both, a canonical and
the isochrone with overshoot.
We note  that the brightness and  color of He-burning  clump stars are
very-well reproduced by the BaSTI isochrones for the adopted reddening
and distance modulus.

The  best fit of  main sequence  using the  Victoria-Regina isochrones
(VandenBerg  et al.\  2006) can  be achieved  with a  distance modulus
equal  to 9.56  mag  and  a reddening  of  0.03.  The  Victoria-Regina
isochrones do not  follow through the phase of  core He burning, hence
we cannot test a match to the red clump stars.

Sandquist  (2004) has  shown  that  a number  of  isochrones from  the
literature are unable to fit the location of MS at $V$ $>$ 15.5.  This
result is  confirmed by  a fit  of Padova (Girardi  et al.\  2000) and
Yonsei-Yale isochrones  (Yi et al.\ 2001)  to our data  shown in Fig.\
\ref{cmd_teo}. As suggested by  Sandquist (2004) the available $T_{\rm
eff}$-color  transformations  are  poor  for main  sequence  stars  at
$T_{\rm  eff}$   $<$  5000  K.   This  deficiency  is   corrected  for
Victoria-Regina   isochrones    by   applying   semi-empirical   color
transformations (VandenBerg  \& Clem 2003).  We note that in  the same
$T_{\rm eff}$ range the theoretical  isochrones do not follow well the
MS stars in metal-poor globular clusters as well (Bedin et al.\ 2001).
This  can largely be  attributed to  the same  problem in  the adopted
$T_{\rm eff}$-color relations.

The  main result  of  our  isochrone fit  to  the new  color-magnitude
diagram  for M~67  is  the  confirmation of  a  small convective  core
overshoot required to  achieve a close match to  the features near the
main sequence turnoff region.
This  result is in  very good  agreement with  the earlier  results by
Sandquist (2004)  and VandenBerg  \& Stetson (2004).  We wish  to note
that in  this study we  have performed comparisons between  the theory
and observations  by employing four independent  and updated libraries
of stellar models.

%
\section{The Catalogue}
\label{CAT}
%

%
\begin{figure*}[!ht]
\centering
\includegraphics[width=\textwidth]{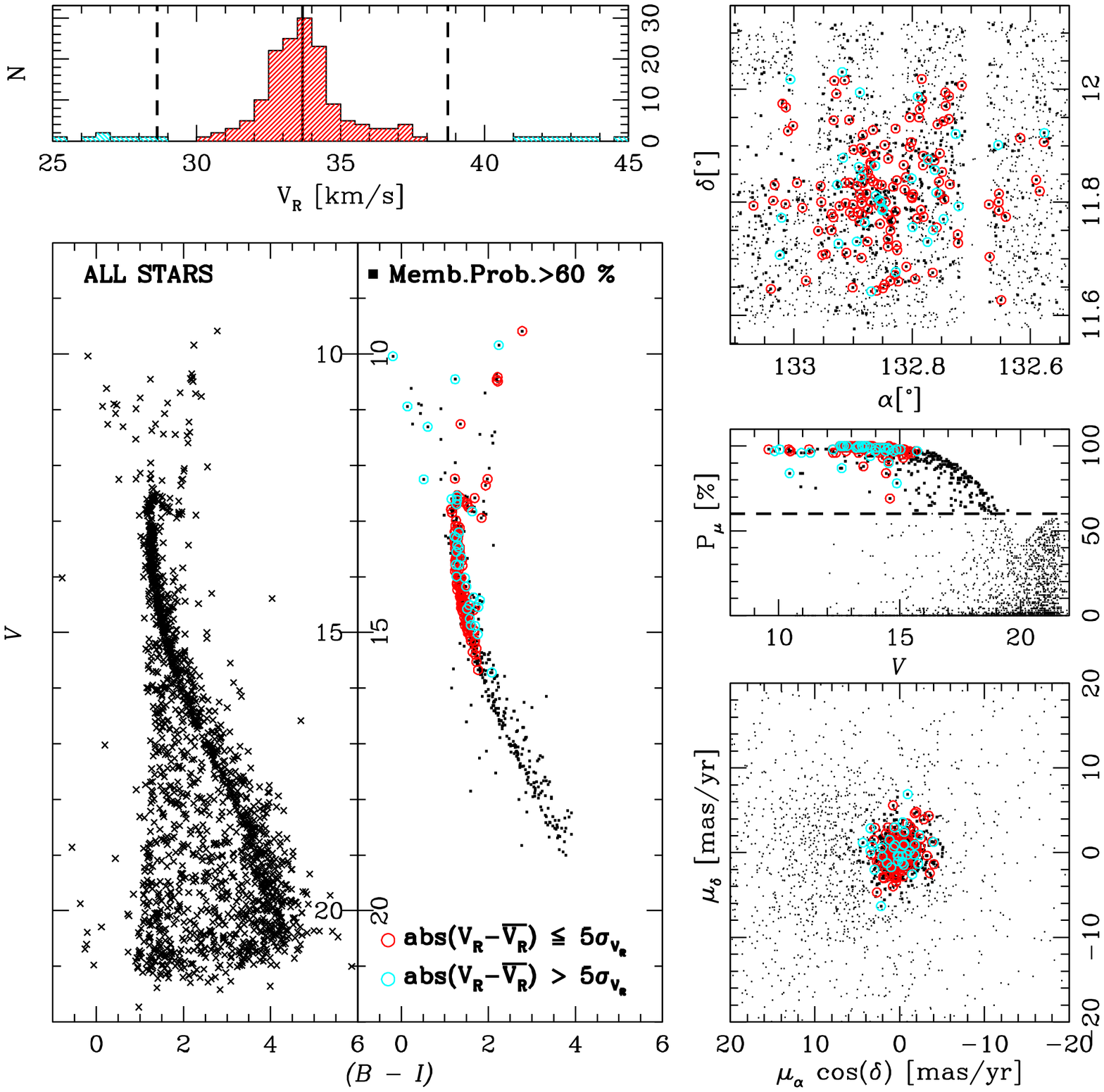}
\caption{
  This plot  is a summary of  the quantities listed  in the catalogue.
  Positions ($\alpha$,  $\delta$), photometry (in $B$,  $V$, and $I$),
  radial  velocities  (V$_{\rm  R}$),  proper  motions  ($\mu_{\alpha}
  \cos{\delta}$,   $\mu_{\delta}$)    and   membership   probabilities
  (P$_\mu$).  Thin dots  are stars with a P$_\mu  \le 60\%$, for thick
  dots P$_\mu > 60\%$. Red circles  are stars for which $\| V_{\rm R}-
  \overline{V_{\rm R }}\| ~ < ~  5 \ \sigma_{V_{\rm R }}$, if not they
  are marked in  cyan. Note that many of  those are potentially, short
  period binaries.
}
\label{summary}
\end{figure*}
%

The catalogue is available electronically at Astronomy \& Astrophysic
(or also under request to the authors).

Description of the catalogue:  column (1) contains the running number;
columns (2) and  (3) give the J2000 equatorial  coordinates in decimal
degrees for the epoch J2000.13,  while columns (4) and (5) provide the
pixel  coordinates $x$  and  $y$ in  a distortion-corrected  reference
frame.  Columns  (6) to  (11)  give  photometric  data, i.e.,  $B,V,I$
magnitudes and their errors.  If  photometry in a specific band is not
available, a flag equal to {\sf  99.999} is set for the magnitude, and
{\sf  0.999} --  for the  error.   Next four  columns (12)-(15),  give
relative proper motions and  their standard errors.  Column (16) gives
the  proper-motion  membership  probability  $P_\mu$.   The  last  two
columns  (17) and  (18) give  radial  velocity (V$_{\rm  R}$) and  its
error, if available, otherwise the table reports {\sf 999.999}.

Figure \ref{summary} shows a summary  of the basic properties of stars
in the  catalogue. Here we  plotted a histogram of  radial velocities,
the spatial  distribution of all stars,  color-magnitude diagrams, the
distribution   of  membership   probabilities,   and  a   vector-point
diagram. They provide additional information over similar plots in the
previous figures.

\begin{sidewaystable*}
\begin{minipage}[t][180mm]{\textwidth}
\caption{First lines of the catalogue available electronically.}
\label{catalog}
\centering
\begin{tabular}{cccccccccccccccccc}
\hline
\hline
     & & & & & & & & & & & & & & & & & \\
  ID & $\alpha$   & $\delta$ & $x$        & $y$ & 
 $B$ & $\sigma_B$ & $V$      & $\sigma_V$ & $I$ & $\sigma_I$ & 
$\mu_\alpha\cos{\delta}$ & $\sigma_{\mu_{\alpha}\cos{\delta}}$ & $\mu_{\delta}$ & $\sigma_{\mu_{\delta}}$ & 
$P_{\mu}$ & $V_{\rm R}$ & $\sigma_{V_{\rm R}}$ \\
 & & & & & & & & & & & & & & & & & \\
(1)     & (2) & (3) & (4) & (5) & (6) & (7) & (8) & (9) & (10) & (11) & (12) & (13) & (14) & (15) & (16) & (17) & (18) \\
     & & & & & & & & & & & & & & & & & \\
%
{[}\#{]} & [ $^\circ$ ] & [ $^\circ$ ]
&[pixel]&[pixel]&
[mag]&[mag]&[mag]&[mag]&[mag]&[mag]&
[mas/yr]&[mas/yr]&[mas/yr]&[mas/yr]&
[$\%$]&[km/s]&[km/s]\\
  & & & & & & & & & & & & & & & & & \\
\hline
 & & & & & & & & & & & & & & & & & \\
%
 1 & 132.897815 & 11.575597 & 3524.156 & -246.966 & 99.999 & 0.999 & 16.064 & 0.009 & 99.999 & 0.999 &    1.73 &  3.51 &    1.73 &  1.25 &  95 &     34.251 &     0.116 \\
 2 & 132.867489 & 11.576494 & 3973.470 & -233.047 & 99.999 & 0.999 & 20.236 & 0.040 & 99.999 & 0.999 &    0.06 &  3.21 &   -1.25 &  3.57 &  39 &    999.999 &   999.999 \\
 3 & 133.092502 & 11.576703 &  639.898 & -230.841 & 99.999 & 0.999 & 21.698 & 0.118 & 99.999 & 0.999 &   13.86 & 36.24 &   -5.12 & 13.92 &  29 &    999.999 &   999.999 \\
 4 & 132.874183 & 11.577474 & 3874.268 & -218.312 & 99.999 & 0.999 & 18.996 & 0.013 & 99.999 & 0.999 &    7.14 &  6.31 &   -7.26 &  5.00 &   2 &    999.999 &   999.999 \\
 5 & 133.093782 & 11.577612 &  620.936 & -217.087 & 99.999 & 0.999 & 21.452 & 0.260 & 99.999 & 0.999 &   22.37 &  7.79 &  -22.55 & 13.39 &   2 &    999.999 &   999.999 \\
 6 & 132.773239 & 11.578593 & 5369.865 & -199.708 & 99.999 & 0.999 & 17.234 & 0.010 & 99.999 & 0.999 &   25.23 &  6.43 &  -29.33 & 28.44 &   0 &    999.999 &   999.999 \\
 7 & 132.614021 & 11.578638 & 7728.562 & -194.667 & 99.999 & 0.999 & 18.647 & 0.008 & 99.999 & 0.999 &    5.41 &  0.24 &   -2.62 &  2.38 &  28 &    999.999 &   999.999 \\
 8 & 133.035261 & 11.579114 & 1487.807 & -194.539 & 99.999 & 0.999 & 20.867 & 0.046 & 99.999 & 0.999 &   15.05 & 24.16 &    5.71 &  2.32 &   7 &    999.999 &   999.999 \\
 9 & 132.949617 & 11.579316 & 2756.606 & -191.207 & 99.999 & 0.999 & 15.411 & 0.009 & 99.999 & 0.999 &   18.39 &  1.25 &    2.32 &  5.06 &   0 &     90.264 &     0.060 \\
10 & 132.829375 & 11.579519 & 4538.135 & -186.755 & 99.999 & 0.999 & 19.515 & 0.042 & 99.999 & 0.999 &    6.31 &  6.19 &  -13.15 & 10.00 &   0 &    999.999 &   999.999 \\
%
 ... &    ...     &     ...    &   ...    &   ...    &  ...  &  ... &  ... &  ... &  ... &  ... &  ... &  ... &  ... &  ...  &  ...  &  ...  &  ... \\
 & & & & & & & & & & & & & & & & & \\
\hline
\label{catalogT}
\end{tabular}
\vfill
\end{minipage}
\end{sidewaystable*}

\section{Conclusions}

In this paper, we applied the techniques developed by Anderson et al.\
(2006,  Paper~I) to a  cornerstone open  cluster M~67.  We demonstrate
that  the CCD  observations taken  just four  years apart  can provide
accurate proper motions, aiding  to decontaminate the cluster CMD down
to  $V$$\sim$20.  An  open  issue   with  the  open  clusters  is  the
intrinsically low  spatial density  of cluster members,  especially in
the  outer  parts  of  clusters,  that makes  it  difficult  to  apply
rigorously the local-transformation approach.  We believed that with a
geometrically  stable CCD  mosaic  (the $WFI@2.2$m  apparently is  not
such) it is possible to  use the global plate solutions for astrometry
and avoid the problem of inadequate local reference frame.

The high precision of our astrometry and $B,V,I$ photometry once again
underscores the  importance of  accurate representation of  PSF across
the entire field-of-view, exemplified  by the concept of empirical PSF
(Paper~I).

Our new photometry confirm that  in M~67 the convective core overshoot
should  be small,  as already  pointed  out by  Sandquist (2004),  and
Vandenberg \& Stetson (2004).

A  substantial addition  to  the  existing data  sets  are our  proper
motions at $V$$>$16,  by up to 5 magnitudes  deeper than the published
proper  motion catalogues.  High-precision radial  velocities  for 211
stars down to  $V=16$ are further helping to establish  a list of bona
fide cluster  members in M~67.  One of the  goals of this paper  is to
share  the   new  data  with  the  astronomical   community  prior  to
comprehensive analysis  and finalizing of cluster  membership at faint
magnitudes.

%
\begin{acknowledgements}
R.~K.~S.~Yadav is grateful to University  of Padova for the Grant that
made possible this Indo-Italian collaboration.
I.~Platais gratefully  acknowledges support from  the National Science
Foundation through grant AST 04-06689 to Johns Hopkins University.
We  thank  the  anonymous  referee  for  careful  reading  and  useful
suggestions.  We also thank Ben Taylor for providing us with their new
unpublished catalogue in electronic format.

\end{acknowledgements}
%

%

%

\end{document}